\documentclass[iop,appendixfloats,revtex4]{emulateapj}
\usepackage{natbib}
\usepackage{hyperref}
\usepackage{amsmath}
\usepackage{graphicx}
\usepackage{epsfig}
\usepackage{xcolor}
\usepackage{ulem}
\usepackage{gensymb}
\usepackage{textcomp}

\shorttitle{Sticking of molecules on ice}
\shortauthors{He et al.}
\begin{document}

\title{Sticking of molecules on non-porous amorphous water ice}

\author{Jiao He\altaffilmark{1}, Kinsuk Acharyya\altaffilmark{2,3}, \& Gianfranco Vidali\altaffilmark{1}}

\affil{$^1$Physics Department, Syracuse University, Syracuse, NY 13244, USA} %
\affil{$^2$PLANEX, Physical Research Laboratory, Ahmedabad, 380009, India (Current Address)}
\affil{$^3$Department of Chemistry, University of Virginia, VA 22904, USA} %
     \email{jhe08@syr.edu, gvidali@syr.edu}

\begin{abstract}
Accurate modeling of physical and chemical processes in the interstellar medium requires detailed knowledge of how atoms and molecule adsorb on dust grains. However, the sticking coefficient, a number between 0 and 1 that measures the first step in the interaction of a particle with a surface, is usually assumed  in simulations of ISM environments to be either 0.5 or 1. Here we report on the determination of the sticking coefficient  of H$_2$, D$_2$, N$_2$, O$_2$, CO, CH$_4$, and CO$_2$ on non-porous amorphous solid water (np-ASW). The sticking coefficient was measured over a wide range of surface temperatures using a highly collimated molecular beam. We showed that the standard way of measuring the sticking coefficient --- the King-Wells method --- leads to the underestimation of trapping events in which there is incomplete energy accommodation of the molecule on the surface. Surface scattering experiments with the use of a pulsed molecular beam are used instead to measure the sticking coefficient. Based on the values of the measured sticking coefficient we suggest a useful general formula of the sticking  coefficient as a function of grain temperature and molecule-surface binding energy. We use this formula in a simulation of ISM gas-grain chemistry to find the effect of sticking on the abundance of key molecules both on grains and in the gas-phase.
\end{abstract}

\section{Introduction}
It is now well recognized that dust grains and ices play an important role in the formation of molecules in the ISM, from the simplest --- H$_2$--- to ones that might be considered to be the building blocks of pre-biotic molecules. Simulation codes now routinely incorporate gas-grain reactions in the study of the chemical evolution of ISM environments \citep[e.g.][]{Garrod2008}. Laboratories have provided a wealth of data on gas-grain interactions to make these simulations realistic, providing binding energies of atoms and molecules to grains,  efficiencies of molecule formation on grains, and the reaction pathways due to the interaction of energizing radiation with ices, to name a few \citep{Vidali2013}. However, little attention has been paid to the process of adsorption of atoms and molecules on grains. Often, it is assumed that at the  temperature of interest in dense cloud environments, where ice-coated grains are at  $\sim $10 K, all gas-phase atoms and molecules are assumed to stick to grain surfaces with either 50\% \citep[e.g.][]{Aikawa2012} or 100\% \citep[e.g.][]{Chang2012} probability, regardless of the temperature of dust grains or of incident molecules. However, this might be an oversimplification. Furthermore, there is now a widespread interest in understanding star formation phases, such as when a young stellar object emits copious radiation that warms up the surrounding medium and causes either desorption of radicals or chemical reactions to occur via diffusion processes. As ices warm up, it is important to know whether certain gas-phase molecules are able to stick to the surface of grains and with what probability.

The sticking of molecular hydrogen on surfaces of grains is of obvious importance, both because H$_2$ is by far the most abundant molecule, and because it has important roles in the physical and chemical evolution of ISM environments. To define sticking, we look first at the process of interaction of a gas-phase particle with a surface. The result of the interaction can be: a) the instantaneous reflection with little or no change in the energy (elastic and inelastic scattering); b)  the  trapping of the particle for a short period of time; or c) the ``permanent'' residence of the particle on the surface. In the third case, the particle essentially becomes thermalized with the surface and will desorb at a rate governed by the temperature of the surface. While the typical technique to study sticking, the King-Wells method (to be described in the Appendix), would measure this third contribution with various degrees of success \citep{Zangwill1988}, the trapping process is also important in astrochemistry. During  trapping, the molecule might sample a larger part of the surface than in the case it were fully thermalized with the surface (for an example of calculations of trapping and sticking, see \citet{Medina2008}). While techniques to distinguish these tapping-sticking events are used in the study of adsorption processes on well characterized crystal surfaces \citep{Rettner1986}, they haven't been employed, to our knowledge,  in astrochemistry. This work is a step in the direction of providing a richer set of data on sticking of molecules on amorphous solid water for simulations and interpretation of observational data.

There have been a few theoretical studies about sticking on surfaces of interest to astrochemistry, mostly about the sticking of H/H$_2$ and isotopologues on amorphous solid water. \citet{Hollenbach1970} in calculating the sticking of H on poly-crystalline water used an expression for the sticking that can be approximated as  $S(T_{\rm g}) \approx (\Gamma ^2 +0.3 \Gamma ^3)/(1 + 2.4 \Gamma + \Gamma^2 + 0.8\Gamma^3)$, where $\Gamma=E_{\rm C}/k_{\rm B}T_{\rm g}$, $T_{\rm g}$ is the gas temperature and  $E_{\rm C}$ is a parameter that depends on the adsorption well depth and the energy transfer in a single collision.
In several other studies, molecular dynamics (MD) has been used to determine the sticking of H/H$_2$ on grain surfaces with different incident beam energies \citep{Masuda1997,Al-Halabi2007,Veeraghattam2014}. The sticking of heavier molecules, such as CO, was also calculated \citep{Al-Halabi2003,Al-Halabi2004} using MD\@. Other than the incident energy, the dust temperature is also affecting the sticking rate. \citet{Cazaux2011} and \citet{Veeraghattam2014} used  classical/quantum dynamics methods to determine the sticking of H/H$_2$ on grain surfaces with selected grain surface temperatures. Compared with theoretical calculations, very limited laboratory measurements are available. Typically, King-Wells and related methods were used to study sticking. \citet{Govers1980} studied the sticking of H$_2$ and D$_2$ on a doped Si slab, and found that the sticking coefficient increases with coverage. \citet{Amiaud2007} measured the sticking of D$_2$ on np-ASW when the surface is at 10~K, and with D$_2$ beam at room temperature. \citet{Matar2010} measured H$_2$ and D$_2$ sticking on non-porous amorphous solid water (np-ASW) with different incident energies at a grain surface temperature of 10 K. \citet{Hornekaer2003} measured D$_2$ sticking on both porous and non-porous ASW\@. Recently, \citet{Acharyya2014} measured the sticking of H$_2$ on olivine substrate by comparing gas load at room temperature and at low temperatures. These King-Wells-like measurements are best suited for sticking measurements at higher surface temperatures, but are less applicable at cryogenic temperatures, as is discussed in the Appendix. In this new work, time-resolved surface scattering measurements are used, which have advantages over the King-Wells method. Other than H$_2$/D$_2$, the sticking coefficient of other molecules at ISM relevant conditions have not been directly measured as far as we know. In this contribution we fill up the gap and measure the sticking coefficient of several other ISM relevant molecules at different surface temperatures.

In gas-grain chemical network models, to use a global sticking coefficient of unity or 0.5 is an over-simplification. (Here the sticking coefficient --- or sticking probability --- is defined as the fraction of incoming particles that become thermalized with the  dust grain surface). Simulations by \citet{Cazaux2011} and \citet{Acharyya2014} have already shown that sticking  plays a role in the chemistry of the ISM\@. It would be useful to have the sticking coefficient for a wider range of radicals and molecules. In this contribution, we analyzed the sticking coefficient ($S$) of both weakly bonded molecules (H$_2$, D$_2$, CO, N$_2$, O$_2$, and CH$_4$) and more strongly bonded molecules (CO$_2$ and NH$_3$) on np-ASW (except for NH$_3$, which was on crystalline water ice) at different surface temperatures ($T_{\rm s}$).   We find that the sticking coefficient of these molecules exhibit similar trends and is a function of $T_{\rm s}$ and of the  binding energy of the molecule with the surface ($E_{\rm des}$). We propose a formula $S(T_{\rm s}, E_{\rm des})$, which can be used to estimate the sticking of other radicals/molecules.  We used this formula in a gas-grain chemical network model to explore the astrophysical implications of these new determinations of the sticking coefficient.

The measured sticking coefficients of N$_2$ and CO may also help to understand the observed anti-correlation between N$_2$H$^+$ and CO\@. Observations and modeling show that the binding energy ratio between N$_2$ and CO needs to be $\sim0.65$ in order to explain the observational results \citep[e.g.]{Bergin1997,Qi2013}. However, various laboratory measurements found that the difference in binding energy between N$_2$ and CO is much higher than 0.65 \citep{Oberg2005,Bisschop2006,Fayolle2016,He2016b}. An explanation of the observed anti-correlation is still lacking. Here we present for the first time a direct comparison of the sticking of N$_2$ and CO to see whether the sticking difference helps to explain the anti-correlation.

The rest of this work is organized as follows: the next section describes the apparatus and how the sticking coefficient was measured using both the King-Wells method and the time-resolved surface scattering method; Section~\ref{sec:result} presents the experimental results, followed by a discussion of how to generalize the measured results; in Section~\ref{sec:implications} we apply the measured sticking coefficients to a gas-grain chemical network model to find out the astrophysical implications; Section~\ref{sec:summary} summarizes this work.

\section{Experimental Setup}
\subsection{Apparatus}
Here we briefly summarize the main features of the apparatus that was used for the measurements of the sticking coefficient. A more detailed description is found elsewhere \citep{He2011,Jing2013,He2015a}. However, the beam modulation and time-of-flight (TOF) measurement methods, which are important for this contribution, are described in detail. The experiments were carried out in a 10 inch diameter ultra-high vacuum chamber (``main chamber''). After bake-out, a pressure as low as $1.5\times10^{-10}$ torr is reached routinely. The operating pressure during measurements is $\sim 3\times 10^{-10}$ torr. A 1~cm$^{2}$ gold coated copper disk substrate is located at the center of the chamber. It can be cooled down to 8~K by liquid helium or heated to 450~K using a cartridge heater. A Lakeshore 336 temperature controller with a calibrated silicon-diode (Lakeshore DT670) is used to measure and control the sample temperature with an uncertainty of less than 0.5~K. A Hiden Analytic triple-pass quadrupole mass spectrometer (QMS) is mounted on a rotary platform to record desorbed/reflected species from the sample or to measure the  composition of the incoming beam. A multichannel scaler (MCS) is coupled with the QMS for in-phase pulse counting. A Teflon cone is attached to the entrance of the QMS detector in order to maximize the collection  of molecules desorbed or scattered from the surface. It has also the function of rejecting molecules  desorbing from other parts of the sample holder. In the King-Wells and surface scattering measurements, the QMS is placed at $42\degree$ from the surface normal, while the incident angle of the molecular beam is $8\degree$ and on the opposite side of the QMS detector. At the back of the sample there is a gas capillary array for water vapor deposition. The capillary array is not directly facing the sample holder in order to obtain a deposition of molecules from the background. Distilled water underwent at least three freeze-pump-thaw cycles to remove dissolved air. A leak valve is used to control the water vapor flow into the main chamber. Because too fast a deposition rate could increase the porosity of ice \citep{Bossa2015}, we regulate the vapor pressure to be $\sim 5\times 10^{-7}$ torr, which corresponds to a deposition rate of 0.5 ML/s. This rate is close to the slowest deposition rate used by \citet{Bossa2015}.   At the temperature (130K) of the substrate during deposition, water vapor forms np-ASW on the substrate \citep{Stevenson1999}. The ice thickness is calculated by integration of the chamber pressure with time, assuming $1\times10^{-6}$ torr$\cdot$s exposure corresponds to 1 monolayer (ML). After deposition, the ice sample is annealed at 130~K for 30 minutes before cooling down for further experiments.

Connected to the main chamber are two highly collimated three-stage molecular beam lines. In this contribution we use only one. The first stage of the beam line houses a Pyrex glass radio frequency (RF) dissociation source. Gas flow to the dissociation source is controlled by an Alicat MCS-5 mass flow controller. The end of the dissociation source is capped by a 1~mm inner diameter aluminum nozzle. The length ($\sim$3~mm) and shape of the nozzle causes a slight collimation on the beam. Therefore the velocity of beam does not follow a perfect Maxwell-Boltzmann distribution (see Section~\ref{sec:direct_beam}). In the second stage of the beam line, there is a chopper wheel with a single 1/40 slit opening and is driven by a DC motor spinning at $50\pm1$ Hz. The corresponding beam pulse width is therefore 0.5~ms. A LED and photodiode pair located at opposite sides of the chopper wheel is used to generate a TTL signal that serves as a trigger signal for in-phase detection. The photodiode pulse signal is half a cycle apart from the beam pulse and triggers a multichannel scaler (MCS) for  time-resolved pulse counting (see Figure~\ref{fig:beam_measurement}). A computer-controlled flag in the third stage controls the exposure time of molecules to the sample.

\subsection{Direct Beam Measurement}
\label{sec:direct_beam}
\begin{figure}
  \epsscale{1}
  \plotone{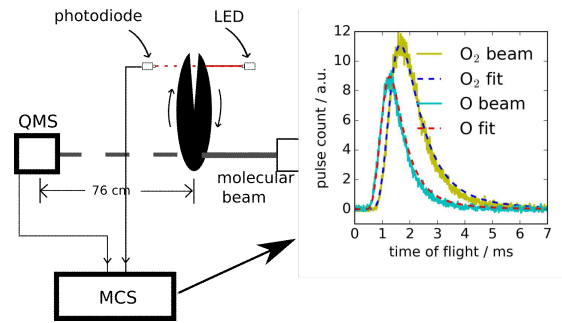}
  \caption{Schematic of the beam measurement. The right side shows the Time-Of-Flight (TOF) measurement of the O$_2$ and O direct beams using the QMS and MCS\@. The fittings were obtained using a Maxwell-Boltzmann distribution assuming a temperature of 300~K. The broadening effect of the 1/40 slit width was taken into account.}
\label{fig:beam_measurement}
\end{figure}

The time-of-flight (TOF) setup was tested in the measurement of a molecular oxygen  beam entering the main chamber. The pulses were recorded by the MCS and were added up for a number of cycles to get a high signal-to-noise ratio. The measured TOF pattern of a O$_2$ beam is shown in Figure~\ref{fig:beam_measurement}. A Maxwell-Boltzmann velocity distribution was used to simulate the TOF signal. After correcting for  beam broadening due to the width of the chopper slit, a good fit was found by setting the temperature to 300~K. This suggests that the beam is effusive and at room temperature. The tail of the fitted curve is slightly higher than the measured one because the nozzle attached to the end of the gas source  has a slight collimating effect. A similar measurement was performed on an atomic oxygen beam (RF power on). The atomic oxygen is also at 300 K, indicating that the atomic oxygen produced by the RF dissociation source is in thermal equilibrium with the Pyrex glass wall.

\subsection{Sticking Measurement: King-Wells Method}
\label{sec:exp-king-wells}
To obtain the sticking probability of  molecules on surfaces, it is customary to use the well-known King-Wells method \citep{King1972}. In the King-Wells measurement, the pressure in the chamber is measured when molecules are introduced. If all molecules are reflected, the pressure is higher than if molecules are adsorbed (stuck) on the surface. In typical measurements, a QMS is  positioned in front of the sample to maximize the detection of molecules reflected from the sample. After the beam is introduced into the main chamber, the QMS measured signal should increase immediately to a value corresponding to the increase of the chamber pressure due to the initial reflection of molecules from the surface. This method suffers from a few technical problems which make its interpretation sometimes difficult. These issues are discussed in the Appendix. Here we mention one limitation of such measurement, specifically the systematic undercounting of sticking events. In the conventional King-Wells method, molecules that are temporarily trapped on the surface are re-emitted in the gas phase after a short time, and therefore are not counted as being stuck. Yet, these molecules --- during their time on the surface --- can diffuse and react with other species on the surface.

\subsection{Sticking Measurement: Time-Resolved Surface Scattering}
\label{sec:scattering}
 We used  time-resolved surface scattering to measure the sticking of molecules on np-ASW\@. After the np-ASW was grown and annealed at 130 K, it was cooled down to the desired temperature for the surface scattering experiments. The beam was chopped in pulses and the reflected molecules were measured by the MCS coupled with the QMS\@. When the surface temperature was high, the sticking was negligible and molecules were reflected directly from the surface without delay. The measured pulse count rate was at the highest. As the surface temperature was lowered, sticking increases, and molecules acquire a residence time that increases with decreasing temperature. This residence time shows up as a delay in the TOF profile. If this delay timescale is much shorter than the chopper period (20 ms), one cannot tell whether the measured signal is due to molecules that scatter directly or is due to molecules with a short residence time. On the other hand, if the delay timescale is much longer than the chopper period, molecules desorbing from the surface appear in the TOF profile as an elevated background. During the measurements, the surface coverage was always below 10\% of a monolayer in all cases. Before the scattering measurement at each temperature, the surface was heated up to desorb adsorbed molecules, therefore ensuring that each scattering experiment always starts from a clean surface.

\section{Result and Analysis}
\label{sec:result}
We measured the sticking of D$_2$, H$_2$, O$_2$, N$_2$, CO, CH$_4$, and CO$_2$ on non-porous amorphous water ice using the technique of time-resolved surface scattering. For some molecules, we also carried out measurements using the King-Wells technique. A comparison of results using the two techniques is given in the Appendix. Here we illustrate how the time-resolved surface scattering measurement was done and analyzed using the example of O$_2$ on np-ASW\@. The measurements of the sticking of the other molecules was done similarly, except for NH$_3$, which was done from crystalline ice, because NH$_3$ reflection  occurs at a temperature that ASW changes into crystalline water ice.

\subsection{\texorpdfstring{O$_2$}{O2} Sticking by Time-Resolved Surface Scattering}
\label{sec:o2scatt}
\begin{figure}
  \epsscale{1}
  \plotone{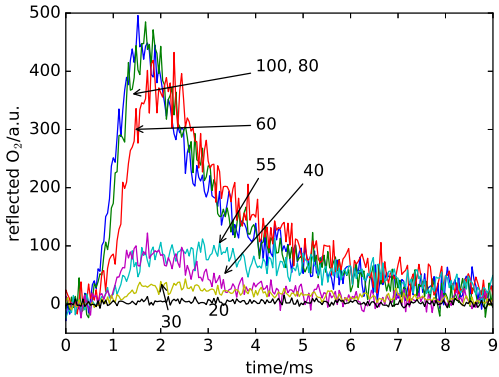}
  \caption{Selected Time of flight (TOF) spectra of O$_2$ in the surface scattering measurement at the surface temperature indicated. Spectra at other temperatures are not shown for clarity. The sample is a non-porous amorphous water ice. The O$_2$ beam was modulated using a 1/40 duty-cycle chopper spinning at 50 Hz.}
\label{fig:o2_scatt_ice}
\end{figure}

TOF spectra of O$_2$ scattering from a np-ASW surface at selected temperatures are shown in Figure~\ref{fig:o2_scatt_ice}. At 120, 100, 90, 80, and 70~K, spectra are similar. This saturation of reflected O$_2$ signal at high temperatures suggests a complete reflection (negligible sticking). Between 60~K and~55 K, the reflection drops dramatically. Below 55~K the reflection decreases slowly to zero. The peak shape at all temperatures are similar except for that at 60~K and 55~K. This change in peak shape at $\sim 55$~K is because of the coincidence of the timescale of O$_2$ residence time at 55~K and the timescale of the TOF peak width. At higher than 60~K, the delay due to the sticking-desorption process is less than the timescale of the peak width, and therefore the molecules that stick cannot be separated from those reflected directly from the surface. The peak width sets the limit of the residence time that can be measured.  At and below 50~K, the residence time scale is longer than the peak width, and the molecules that stick and desorb show up in the TOF spectrum as elevated background.

The reflection rate $R(T_{\rm s})$ is calculated by integrating the peak area of the reflected signal (the TOF peak). It should be noted that at $\sim$55 K, there is some uncertainty introduced by the coincidence of the two timescales as discussed above. $R(T_{\rm s})$ is normalized to be $R_n(T_{\rm s})$. The normalized reflection of O$_2$ from np-ASW is shown in Appendix (Figure~\ref{fig:o2_reflec}). The sticking  $S(T_{\rm s})$ is calculated as $S(T_{\rm s})=1-R_n(T_{\rm s})$.

\subsection{Time-Resolved Surface Scattering of \texorpdfstring{H$_2$, D$_2$, N$_2$, CO, CH$_4$, NH$_3$, and CO$_2$}{H2, D2, N2, CO, CH4, NH3, and CO2}}
\label{sec:n2_co_h2_d2}
\begin{figure}
  \epsscale{1}
  \plotone{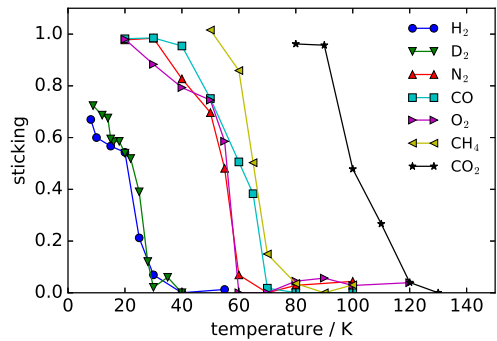}
  \caption{Sticking coefficient of H$_2$, D$_2$, N$_2$, CO,  O$_2$, CH$_4$, and CO$_2$ on np-ASW at different surface temperatures. Lines are guides to the eye.}
\label{fig:sticking_summary}
\end{figure}

Surface scattering experiments were also carried out for  H$_2$, D$_2$, N$_2$, CO, CH$_4$, and CO$_2$ from np-ASW  and NH$_3$ from crystalline ice (CI). The sticking of NH$_3$ from CI is measured up to 155~K at which water ice desorbs significantly, but still no reflected NH$_3$ was measured by the MCS\@. The sticking of NH$_3$ is unity on water ice at least for surface temperature below 155 K. The sticking probability versus temperature for other molecules on np-ASW is shown in Figure~\ref{fig:sticking_summary}. The trends of sticking are similar for N$_2$, CO, O$_2$ and CH$_4$, but there is a slight temperature shift from one to the other. The sticking coefficients presented here are only lower limits of sticking in dense clouds, because the gas temperature in dense clouds conditions is usually lower than in the current study and higher sticking is expected in dense clouds.

\begin{figure}
  \epsscale{1}
  \plotone{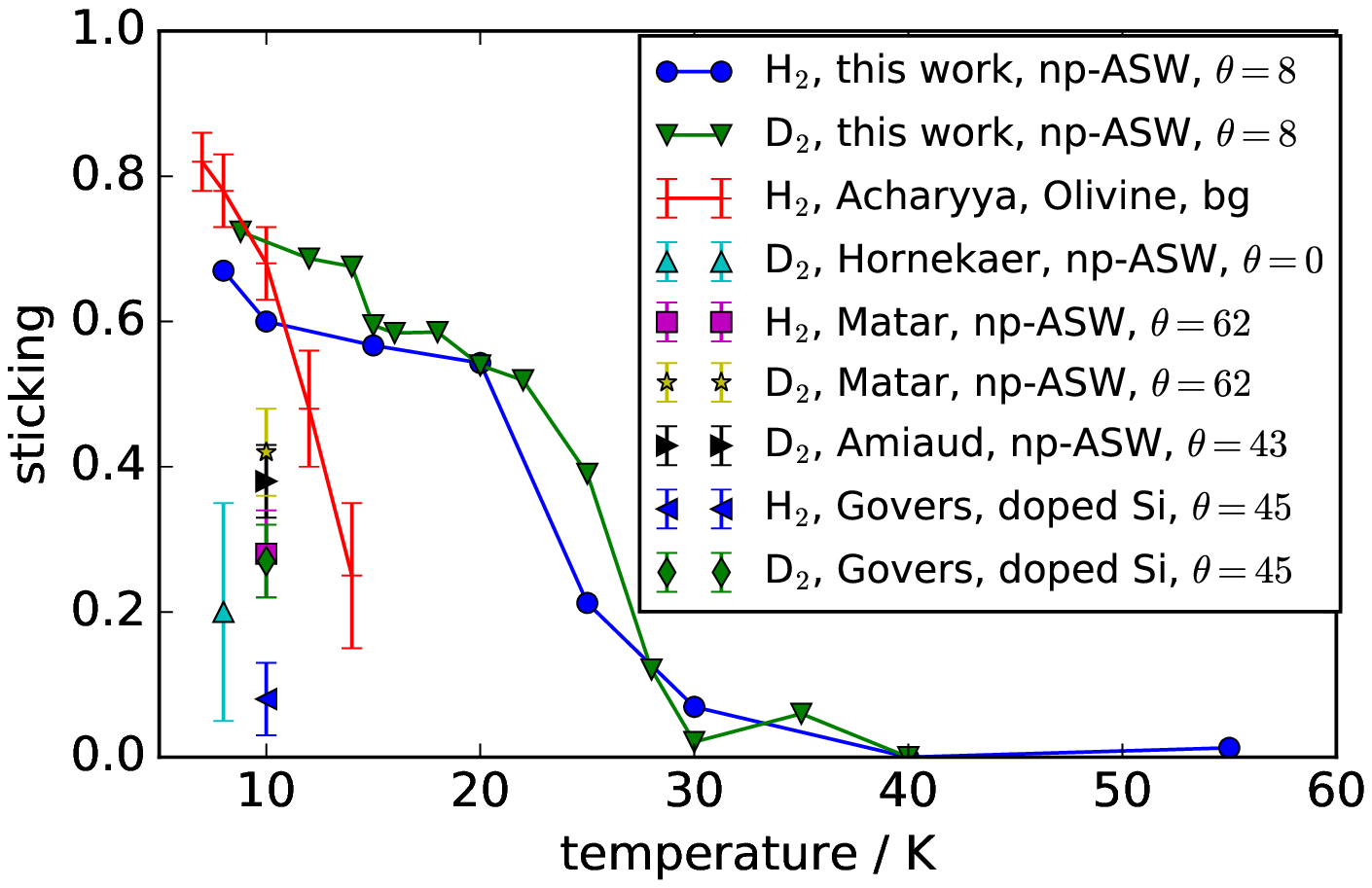}
  \caption{Sticking coefficient of H$_2$, D$_2$ vs.\ sample temperature measured in this study, compared with those by \citet{Acharyya2014}, \citet{Matar2010}, \citet{Hornekaer2003}, \citet{Amiaud2007}, and \citet{Govers1980}. The data points of Matar et al. for H$_2$ and of Govers et al. for D$_2$ overlap. The substrates been used and the angles between the incident H$_2$/D$_2$ beam and the surface normal are indicated in the legend. In \citet{Acharyya2014} H$_2$  was deposited from the background.}
\label{fig:compare}
\end{figure}

We compare the sticking coefficients obtained from this study with those in the literature. The  sticking coefficient of H$_2$ and D$_2$ vs.\ sample temperature obtained in this study is plotted in Figure~\ref{fig:compare} along with the values obtained by others~\citep{Acharyya2014, Matar2010, Hornekaer2003,Amiaud2007, Govers1980}. For these studies, np-ASW is used as the sample surface except in Acharyya (Olivine) and Govers (doped Si slab). The deposition angles are also different; the question is then whether and to what extent it might affect the sticking coefficient \citep{Batista2005}. On clean surfaces, the sticking scales with the square of the momentum perpendicular to the surface, but on disordered surfaces sticking is independent on the incident angle. The sticking of H$_2$ at $\sim 10$~K obtained in this study is close to that obtained by \citet{Acharyya2014}, but at higher sample temperatures, the difference becomes large. At all temperatures, the sticking coefficient from this study is significantly higher than previous studies (except for Acharyya). This is because the conventional King-Wells method used by others over-estimated the reflection events by counting those with short residence time as direct reflection.

\section{Discussion}
In this contribution we measured the surface temperature dependent sticking probability of several key molecules on np-ASW surfaces. These sticking  coefficient at different temperatures can be used in gas-grain chemical network models. However, among the hundreds of species in a typical gas-grain chemical network model, only for a very small fraction the sticking coefficient has been  measured. To get an estimate of the sticking for other species, we came up with an empirical expression for the sticking coefficient as a function of surface temperature. From Figure~\ref{fig:sticking_summary} it appears that the temperature at which the sticking coefficient changes dramatically is correlated with how strongly a molecule binds to the surface, i.e., the binding energy of the molecule with the surface. Therefore we assume that the sticking rate is a function of both surface temperature and binding energy. Ideally, the expression should also take into account the gas temperature. However, because of experimental limitations, this piece of information is unavailable.

\begin{figure}
  \epsscale{1}
  \plotone{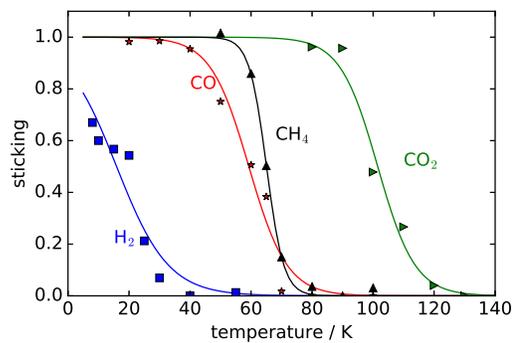}
  \caption{Fitting of the sticking coefficient of H$_2$, CO, CH$_4$, and CO$_2$ as a function of temperature on non-porous water ice using the empirical expression $S= 0.5(1 - \tanh (\beta (T_{\rm s} - \gamma E_{\rm{LC}} ) ))$, where fitting parameters of each molecule are given in Table~\ref{tab:energy}.  Traces of D$_2$, O$_2$ and N$_2$ are omitted for clarity.}
\label{fig:sticking_fit}
\end{figure}

The following expression
\begin{equation}
S(T_s, E_{\rm{LC}, i})= \alpha(1 - \tanh (\beta (T_s - \gamma E_{\rm{LC}, i} ) ))
\label{fit}
\end{equation}
was found to fit the data well. $\alpha$, $\beta$ and $\gamma$ are three fitting parameters. $\alpha$ is set to be 0.5 to ensure that the sticking is ranged between 0 and 1. The best fitting $\beta$ and $\gamma$ values are shown in Table~\ref{tab:energy}. E$_{\rm{LC}, i}$ is the binding energy of a given species $i$ at low surface coverages ($\sim$ 0.001 ML). The binding energies are obtained from thermal programmed desorption (TPD) experiments \citep{He2016b} except for H$_2$ and D$_2$, which are adapted from \citet{Katz1999} and \citet{He2014}, respectively. (The binding energy is the same as the desorption energy if desorption is not activated, which is the case here). Because most of the molecules are present in ISM ices in small quantities with respect to water, it is important to use binding energies that are obtained in experiments where the coverage of the adsorbate on water ice is a small fraction of one layer \citep{He2016b}. For this reason, we carried out all of the surface scattering measurements at low surface coverages. As shown in Table~\ref{tab:energy}, the binding energy of a molecule on np-ASW depends greatly on whether it is obtained in an experiment at  low (sub-monolayer) or high (monolayer) coverage. For the CO$_2$ binding energy, there is no distinction between E$_{\rm ML}$ and E$_{\rm LC}$ because CO$_2$ forms clusters on np-ASW even at very low coverage \citep{He2016b}. In Figure~\ref{fig:sticking_fit} we only present H$_2$, CO, CH$_4$, and CO$_2$. The fitting of other molecules are similar.

\begin{table}
\centering
\caption{Fitting parameters for different molecules in Equation (\ref{fit}), and binding energies for various species on non-porous amorphous water ice. $E_{\rm ML}$ and $E_{\rm LC}$ are binding energies at monolayer coverage and low coverage, respectively. The E$_{\rm LC}$ values for H$_2$ and D$_2$ are taken from \citet{Katz1999} and \citet{He2014}, respectively. The rest are taken from \citet{He2016b}.}
\label{tab:energy}
\begin{tabular*}{0.5\textwidth}{@{\extracolsep{\fill} }ccccccl}
       & $\beta$ & $\gamma$ & E$_{\rm LC}$ (K)& $E_{\rm ML}$ (K) & E$_{\rm LC}$/E$_{\rm ML}$ &  \\
       & \\ \cline{1-6}
    &\\
H$_2$  & 0.059 & 0.051 & 315  & ---   & ---   \\
D$_2$  & 0.072 & 0.029  & 650  &  ---  & ---  \\
N$_2$  & 0.12  & 0.043 & 1250 & 790  & 1.58 \\
CO     & 0.08  & 0.04  & 1480 & 870  & 1.70 \\
O$_2$  & 0.17  & 0.042 & 1310 & 920  & 1.42 \\
CH$_4$ & 0.18  & 0.045 & 1440 & 1100 & 1.31 \\
CO$_2$ & 0.082 & 0.044 & 2320 & 2320 & 1.0
   & \\ \cline{1-6}
   \\
average& 0.11 & 0.042 & --- & --- & ---

\end{tabular*}
\end{table}

\section{Astrophysical Implications}
\label{sec:implications}
Many of the more than 200 molecules detected in the ISM are believed to be formed on the surface of interstellar grains, including molecular hydrogen which is the simplest and the most abundant molecule in the ISM\@. The accretion rate of a given neutral gas phase species $i$ on  a grain is given by,
\begin{equation}
R_{acc}(i)=S_i \sigma v_i n_i n_{\rm d}.
\end{equation}
where, $S_i$ is the sticking coefficient, $\sigma$ is the grain cross-section, $v_i$, $n_i$ are the thermal velocity and concentration of the incoming species, respectively, and $n_{\rm d}$ is the grain number density. Thus sticking of gas-phase species on to the grain surface has a profound effect in the synthesis of complex molecules since it controls the availability of atoms/molecules on the grain surface. However, the change of sticking coefficient with dust grain temperature was poorly constrained.

In this work we used Equation~(\ref{fit}) to represent the laboratory measured sticking coefficient of molecules on np-ASW\@.  For H$_2$, D$_2$, CO, O$_2$, N$_2$, CH$_4$ and CO$_2$ we used the laboratory measured E$_{\rm LC}$, and best fitting $\beta$, and $\gamma$ values in Table~\ref{tab:energy}. To apply the sticking formula to a wider range of molecules of which sticking coefficients are unavailable, we use $\beta=0.11$ and $\gamma=0.042$, which are the average of the values in Table~\ref{tab:energy}, as an estimation. For molecules of which  only $E_{\rm{ML}}$ is available, one can estimate $E_{\rm LC}$ based on the measured species in the table. The ratio $E_{\rm LC}$/$E_{\rm ML}$ varies from $\sim 1.3$ to $\sim 1.7$, except for molecules that form clusters (CO$_2$). We use the median $E_{\rm LC}$/$E_{\rm ML}$=1.5 in the estimation. We utilized the gas-grain chemical network as described in \citet{Acharyya2015a, Acharyya2015b}.  Both the gas-phase and grain surface chemistry are treated via a rate equation approach. The major features of the simulation are as follows:
\begin{itemize}
\item We considerred the so-called classical dust grains having a size of 0.1 $\mu$m with a surface site density $n_{\rm s}$ = 1.5 $\times$ 10$^{15}$ cm$^{-2}$ \citep{Hasegawa1992}, leading to about 10$^6$ binding sites of adsorption per grain.

\item We used standard low-metal elemental abundances, initially in the form of gaseous atoms --- except for molecular hydrogen. Elements having ionization potentials lower than 13.6 eV are in the form of singly charged positive ions, i.e., C$^+$, Fe$^+$, Na$^+$, Mg$^+$, S$^+$, Si$^+$, and Cl$^+$.

\item Physical parameters --- except for the gas and dust temperatures --- remain constant and homogeneous throughout the chemical evolution with a proton density $n_{\rm H}$ = $2 n(\rm H_{2}) + n(\rm H) $ of 2 $\times$ 10$^{4}$ cm$^{-3}$, visual extinction $A_{\rm V}$= 10 mag and standard cosmic ray radiation flux of 10$^{-17}$ s$^{-1}$.

\item We used a warm-up model as described in \citet{Garrod2008}. In the first phase, both the gas and grain temperatures are kept constant at 10 K up to 10$^6$ years; then they are linearly increased to 200 K in 5 $\times$ 10$^4$ years and then kept constant at 200 K till 10$^7$ years. The reason for using a warm-up model is to study the effect of the temperature dependence of sticking on ISM chemistry in hot cores and hot corinos.

\item The binding energy for desorption and hopping is taken from \citet{Garrod2008}. The diffusion-to-binding energy ratio is a key parameter for the study of grain surface chemistry although poorly constrained. \citet{Katz1999} found this ratio to be $\sim$ 0.8 for hydrogen by fitting laboratory data. In absence of laboratory data for other species, several values are used in the literature; $0.3$ \citep{Hasegawa1992}, $0.5$ \citep{Garrod2006, Garrod2008, Acharyya2011}, and a time dependent value \citep{Garrod2011}. We used a value of 0.5 for the diffusion-to-binding energy ratio in accordance with~\cite{Garrod2008}.
\end{itemize}

\subsection{Simulation Results}

\begin{figure}
  \epsscale{1.1}
  \plotone{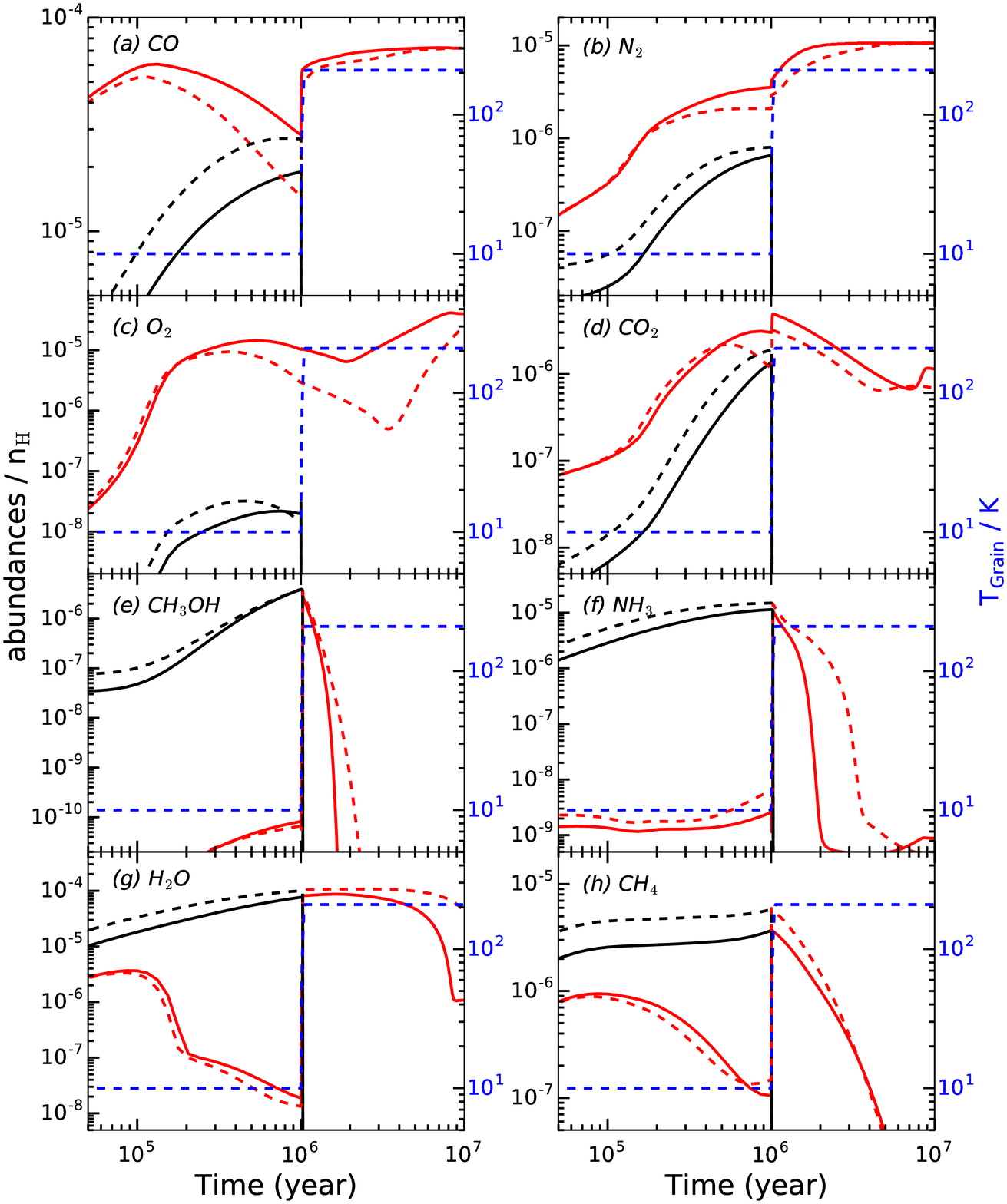}
\caption{ Gas (red) and grain (black) surface abundance of assorted molecules vs.\ time. Solid and dashed lines are for model M1 (sticking coefficient  0.5) and model M2 (sticking coefficient from Equation~(\ref{fit})), respectively. The dashed blue line shows the grain/gas temperature.}
\label{fig-astro1}
\end{figure}

\begin{figure}
  \epsscale{1.1}
  \plotone{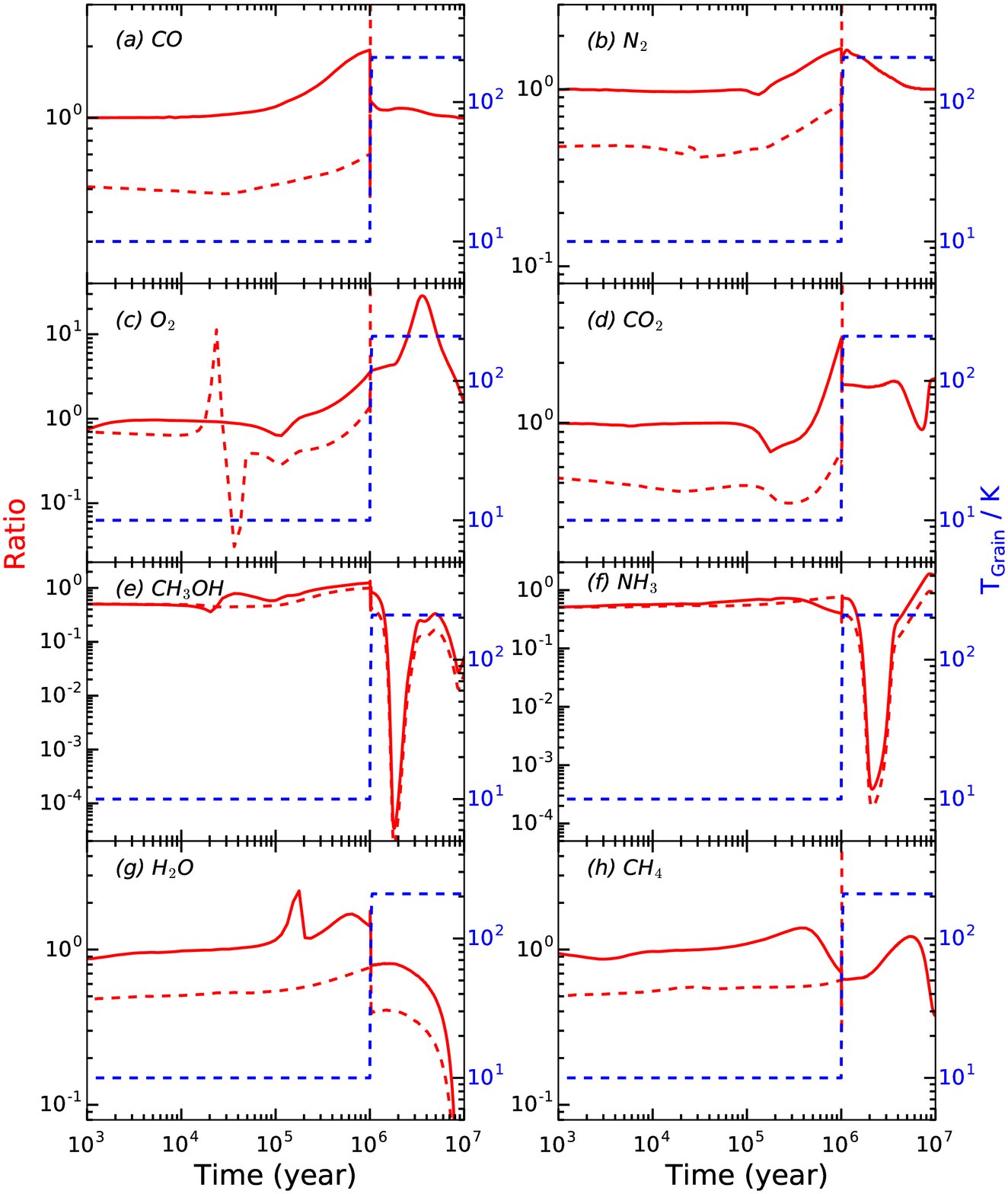}
  \caption{ R$_X$ (ratio between abundances using sticking of 0.5 vs.\ using sticking from Equation~(\ref{fit})) for selected molecules. The solid (dashed) line is used for gas-phase (grain surface) species. The dashed blue line shows the grain/gas temperature. }
\label{fig-astro2}
\end{figure}

Figure~\ref{fig-astro1} shows the change over time of the abundance of selected molecules in the gas-phase
(red traces) and on grains (black). Solid and dashed lines are used when the sticking coefficient is set to
0.5 (Model M1) and by the fitting formula Equation~(\ref{fit}) (Model M2), respectively. Finally, the dashed blue
line shows the variation of grain/gas temperature. We now define a parameter called R$_X$ as follows:
\begin{equation}
R_X= n_{X, \rm M1}/n_{X, \rm M2}
\end{equation}
where, $n_{X, \rm M1}$ and $n_{X, \rm M2}$ are the abundance of a given species $X$ for models M1 and M2,
respectively. Figure~\ref{fig-astro2}  shows  R$_X$ vs.\ time for selected species.

\begin{figure}
 \epsscale{1.1}
  \plotone{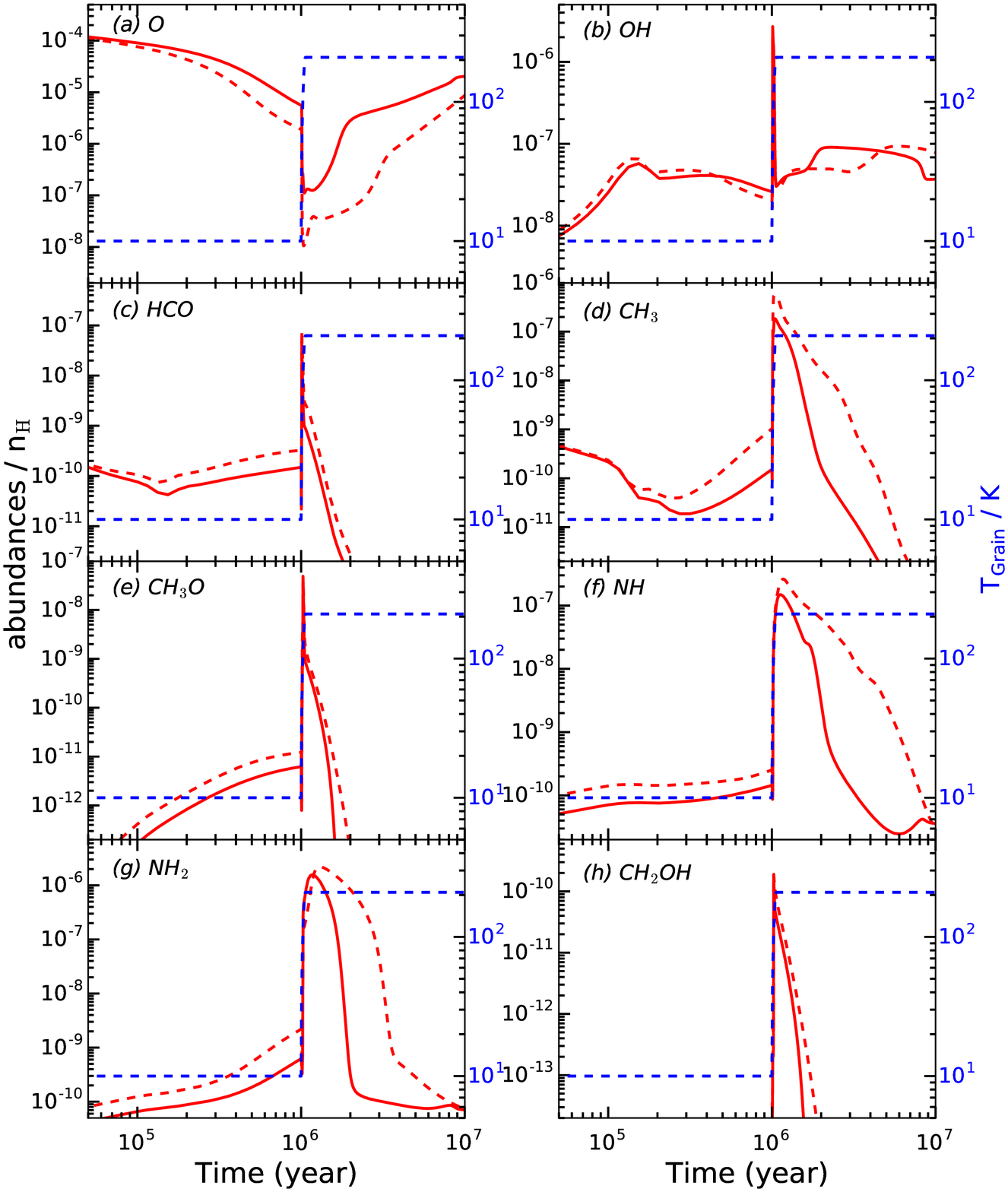}
  \caption{ Gas-phase abundance of selected radicals vs.\ time. Solid and dashed lines are for model M1 (sticking coefficient  0.5) and model M2 (sticking coefficient from Equation~(\ref{fit})), respectively. The dashed blue line shows the grain/gas temperature. }
\label{fig-astro3}
\end{figure}

Figure~\ref{fig-astro1}a shows the abundance  of CO vs.\ time. It is clear that with time the gas-phase CO
depletion becomes much stronger, since at 10 K the sticking coefficient is close to unity
(Figure~\ref{fig:sticking_summary}). This makes grain surface CO to be more abundant ($\sim$ 2) in M2 than in M1
 (Figure~\ref{fig-astro2}),
which in turn increases the abundance of species that have CO as one of their reactants. In a model of cold
cores this effect will be even more pronounced since the grain temperature remains at 10 K up to the end of
the simulation ($\sim$ 10$^7$ years). However, in the models the gas and grain temperatures begin to rise
linearly (warm-up phase) after 10$^6$ years; therefore, CO comes off  the grain surface. The final abundance
after the warm-up phase is slightly lower in M2, the model with the fitting formula.

Figure~\ref{fig-astro1}b shows the abundance of N$_2$ vs.\ time. Like CO, it has a higher depletion in M2.
Since N$_2$ is not directly observable in the gas phase, its abundance is traditionally estimated from the
chemical daughter products N$_2$H$^+$ and NH$_3$. It is observed that CO and N$_2$H$^+$ are anti-correlated,
which is often explained by pointing out that there is a factor of 0.65 difference in the binding energies
of CO and N$_2$ on ASW \citep{Bergin1997}.  However, using our data, the binding energy ratios between
N$_2$ and CO in the monolayer and low coverage regime are $\sim0.91$ and $\sim0.84$, respectively
(Table~\ref{tab:energy}). This is clearly much higher than the value necessary to explain the anti-correlation
of CO and N$_2$H$^+$. Moreover, these molecules have similar sticking coefficient; therefore the anti-correlation
cannot be attributed to either differential sticking or binding energy differences.

The abundance variation of O$_2$ vs.\ time (Figure~\ref{fig-astro1}c) shows a very similar behavior as CO
and N$_2$, i.e., it has higher depletion in model M2, although the grain surface O$_2$ abundance is significantly
lower. Thus, qualitatively CO, N$_2$ and O$_2$ behave very similarly. R$_{\rm O_2}$ is as high as 30 which
could be seen in Figure~\ref{fig-astro2}.
The abundance on grains becomes very low beyond a certain temperature due to rapid thermal desorption;
therefore, the value of  R$_X$ beyond this stage is not very informative because it represents a ratio
between two very small numbers.

Figure~\ref{fig-astro1}d, shows the abundance variation of CO$_2$. Initially, both M1 and M2 have similar
gas-phase abundances, but for times longer than  5 $\times$ 10$^5$ years model M1 has significantly more
CO$_2$ in the gas-phase than model M2 due to less depletion. The grain surface CO$_2$ abundance is always
higher in model M2. Since on the dust grain CO$_2$ is produced via the reaction CO + OH, the availability
of more CO in the model M2 can produce more CO$_2$ on the grain surface. However owing to the very low grain
temperature (10 K) it will not be a significant contribution in the simulation  unless the method of
\citet{Garrod2008} is employed. It could be seen from the Figure~\ref{fig-astro2} that the peak gas
R$_{\rm{CO}_2}$ is close to 3. Figure~\ref{fig-astro1}e shows the methanol abundance; it can be seen that
with time both models M1 and M2 tend to merge. During the warm-up and subsequent phase, the difference becomes
much smaller.  After the warm-up phase, the  methanol abundance in the gas-phase goes down quickly. The ratio
of gas-phase and grain surface methanol  goes down significantly, but this is not particularly significant
because at this time the methanol abundance is low.

Figure~\ref{fig-astro1}f shows the time evolution of  ammonia. The difference between the M1 and M2 models is
significant after the warm -up phase. R$_{\rm {NH}_3}$ can be as low as 2 $\times$ 10$^{-4}$ (Figure~\ref{fig-astro2}),
i.e., when the laboratory measured sticking coefficient is used (model M2) we observe a significantly higher
amount of NH$_3$. A careful look at the reactions involving ammonia reveals that between 10$^6$ years and
5 $\times$ 10$^6$ years, a major source of grain surface ammonia is accretion from the gas-phase. Due to higher
sticking, the amount of ammonia on grain surfaces is higher in model M2 than in model M1. However a fraction
of this ammonia also comes back to the gas-phase via desorption due to the high grain temperature; when in
the gas-phase, it gets destroyed by positive ions. But the net effect is that freezing of more ammonia in
model M2 than model M1 delays its destruction in the gas phase. This could be an efficient process to delay
the destruction of more complex molecules (the ones with binding energy greater than 5000 K).
Figure~\ref{fig-astro1}g shows the abundance of water vs.\ time. It is clear that up to $\sim$ 10$^6$ years,
the grain surface water abundance is a factor of 2 higher in model M2 (Figure~\ref{fig-astro2}) whereas gas-phase
water abundance is almost up to $\sim$ 10$^5$ years; after that, it is generally lower in M2 up to the warm-up
phase due to higher
depletion. At the late time ($\ge 10^6$ years) the difference once again goes up; now in M2 the water abundance
is a factor 2 to 5 higher than in M1 (Figure~\ref{fig-astro2}). Thus, the behavior is similar to the one of
ammonia and the result is that water have significantly higher abundance in the gas-phase for a longer time.
Figure~\ref{fig-astro1}h shows that CH$_4$ on the grain surface in model M2 is a factor 2--3 more abundant than
in M1 up to 10$^6$ years (Figure~\ref{fig-astro2}).
However, for the gas-phase the difference is relatively low and after 3 $\times$ 10$^6$ years there is almost
no difference in abundance between the M1 and M2 models.

Figure~\ref{fig-astro3} shows the abundance of selected radicals vs.\ time. \citet{Garrod2008} argued that radicals such as OH, CO, HCO, CH$_3$, CH$_3$O, CH$_2$OH, NH, and NH$_2$ produced pre-dominantly by cosmic ray-induced photodissociation of the ices during the cold earlier stage of evolution can play a significant role during the warm-up and post warm-up phase in the synthesis of more complex molecules. It is clear from the Figure~\ref{fig-astro3} that only O has lower abundance in model M2. All the other radicals have higher abundance. CH$_3$, NH and NH$_2$ have significantly higher abundance in model M2 which is likely to play a role in the synthesis of more complex molecules involving these species. Further analysis will be presented in a future publication.

\section{Summary}
\label{sec:summary}
We measured the sticking  of H$_2$, D$_2$, N$_2$, O$_2$, CO, CH$_4$, and CO$_2$ on non-porous amorphous solid water as a function of sample temperature. We found that to a good approximation the sticking coefficient  can be expressed as a function $S(T_s,  E_{\rm{LC}}$) of the biding energy of the molecule  at low coverage E$_{\rm{LC}}$ and the the temperature T$_s$ of the surface. We then used this formula in a simulation code to study the time evolution of the chemical make-up of a dense cloud. The results of the simulation using the new data are compared to the results of the simulation using the standard value for the sticking coefficient, 0.5. We find that CO, N$_2$, and O$_2$ abundances vs.\ time behave similarly. There is a greater gas-phase depletion of these molecules in the new (M2) model than in the traditional (M1). The enhanced freezing of ammonia on grains in M2 leads to a delay in its destruction when released in the gas-phase in the warm-up period. A similar trend is observed for water. Finally, most radicals (especially CH$_3$, NH and NH$_2$) required for the synthesis of complex organic molecules during the hot corino phase have enhanced abundance in the model with the sticking formula.

\section{Acknowledgments}
We would like to thank S M Emtiaz and Xixin Liang for technical assistance. This work was supported in part by a grant
to GV from  NSF --- Astronomy \& Astrophysics Division (\#1311958). K. A. like to thank the support of local funds
from Physical Research Laboratory and the University of Virginia and funds from the National Science Foundation
for the program of astrochemistry at the University of Virginia under the direction of Eric Herbst.

\section{Appendix}

In the King-Wells measurement, a molecular beam is aimed at the sample surface. The amount of pressure rise in the sample chamber depends on the fraction of the beam that is reflected from the sample surface. If all of the beam hits the target and all sticks, there is no pressure rise. Conversely if the target is kept at high temperature, then all the beam is reflected giving the maximum pressure increase. Below we illustrate how the sticking coefficient is obtained using the example of O$_2$ on np-ASW\@.

\begin{figure*}
  \epsscale{0.9}
  \plottwo{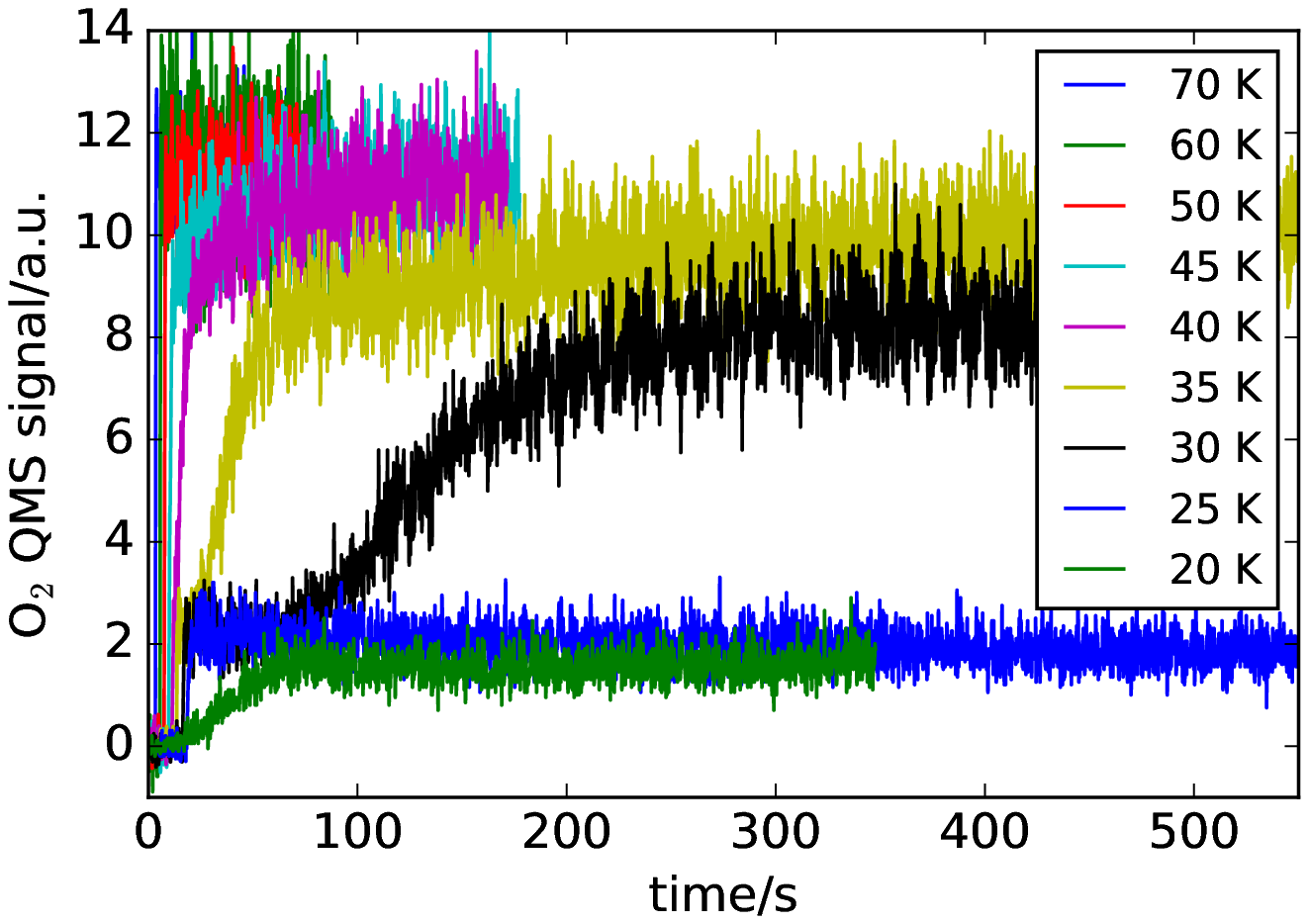}{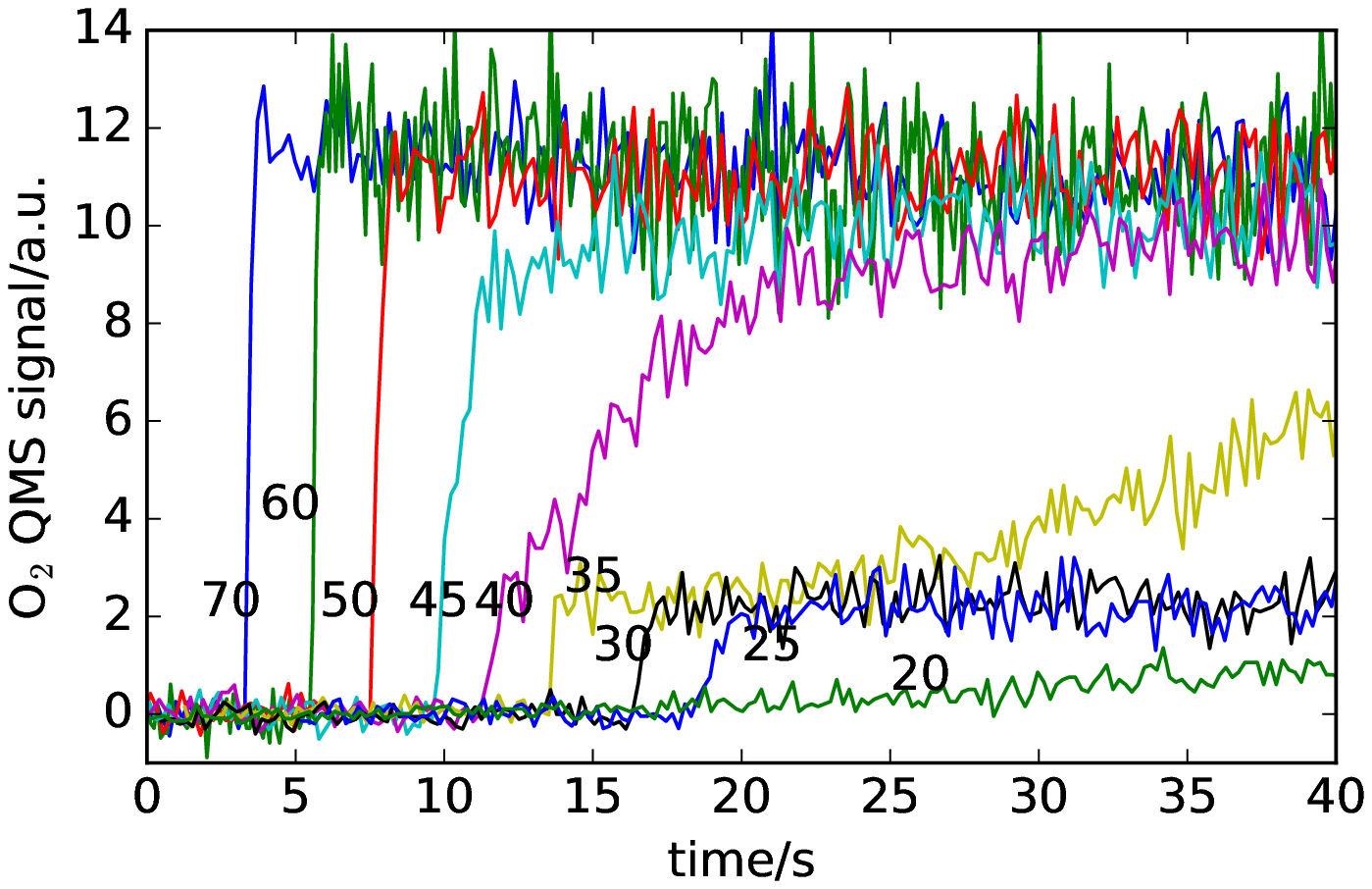}
  \caption{O$_2$ King-Wells measurement at different sample temperatures. In the left panel the whole curve is shown. At each temperature, the O$_2$ signal is measured by the QMS until a saturation level is reached. The right panel only shows the O$_2$ signal right after the beam is introduced in the main chamber. The sample is np-ASW.}
\label{fig:king-wells}
\end{figure*}

The sticking probability $S$ can be calculated based on Figure~\ref{fig:king-wells}.  The left panel shows the whole reflection curve until the saturation level is reached at each temperature, while the right panel zooms in the begining the begining part and shows the O$_2$ signal right after the beam is introduced. At a surface temperature of 70 K, 60 K, and 50 K, after the beam is introduced the O$_2$ signal rises up from the background level to the saturated value immediately. This implies that at 70 K and 60 K, the sticking is zero or that the residence time is shorter than the QMS measurement time scale used in the experiment (longer than the QMS dwell time-- 0.2 s). At 50 K, the O$_2$ saturation level  is slightly lower than that at 70 K or 60 K. This  difference is likely due to the pumping effect of the sample holder. Liquid helium cools down the sample as well as the whole sample holder which acts as a cryogenic pump. As the temperature of the sample is lowered, certain regions of the sample holder may lower to a temperature which is lower than the O$_2$ desorption temperature and the effective pumping speed increases. This may explain why the saturated O$_2$ signal at 50 K is slightly lower than that at 60 K or 70 K. As the surface temperature is lowered further, the saturation level also drops. When the surface temperature is at 25 K or 20 K, there is no saturation of O$_2$ on the sample because O$_2$ ice layers  build up. However, at 25 K and 20 K the  O$_2$ signal seems to increase after the beam is introduced, until a maximum is reached. This is probably because part of the sample holder (not necessarily at 25 or 20~K) is saturated, and there is decrease in effective pumping speed. The gradual saturation of O$_2$ at 45 K, 40 K, and 30 K may also be affected by the change in effective pumping speed. The O$_2$ signal measured by the QMS using the King-Wells method is shown in Figure~\ref{fig:king-wells}.

\begin{figure}
  \epsscale{0.9}
  \plotone{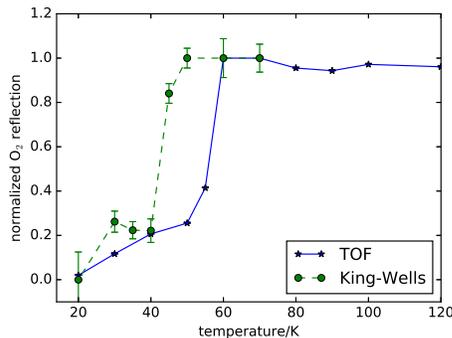}
  \caption{Normalized reflection of O$_2$ measured by the King-Wells method and surface scattering. }
\label{fig:o2_reflec}
\end{figure}
\label{sec:result-king-wells}

Now we summarize the drawbacks of the King-Wells method. The King-Wells method is widely used in surface science to study the sticking of molecules on reactive surfaces, typically metals. A molecule can chemisorb (i.e., it forms a strong bond) on the metal surface, and in this case there is a well defined saturation of the signal. In addition, the temperature range of interest is usually higher than room temperature. In comparison, the sticking of molecules on cosmic dust grains is due to weak dispersion forces, and the interesting temperature range is close to the temperature at which ice layers  build up. This gives the first drawback: the saturation may not be reached (for example, O$_2$ on surface at below 28 K). The second drawback is that the change in pumping speed can affect the reflection result.  Parts of the apparatus  at low temperature act as cryogenic pump with a undefined or changing pumping speed. The third drawback is that King-Wells method does not have enough time resolution to distinguish molecules that reflect directly from molecules that stick but have a relatively short residence time. This lack of time resolution causes overestimation of the reflection rate, and a corresponding underestimation of the sticking rate.

A comparison of the sticking measured by King-Wells method and time-resolved surface scattering method is shown in Figure~\ref{fig:o2_reflec}. The most significant difference between surface scattering and King-Wells method is between 40 K and 60 K. This difference arises because the time resolution of the King-Wells method is of the order of one second in our set-up, while the time resolution for surface scattering is of the order of a few milliseconds. The King-Wells method overestimates the reflection by counting as direct reflection  molecules that stick but have a residence time shorter than $\sim1$ s.
%

\end{document}